\documentclass[preprint,amsmath,amssymb,aps,showpacs,showkeys,superscriptaddress,pra]{revtex4-1}
\usepackage{graphicx}
\usepackage{dcolumn}
\usepackage{bm}
\usepackage{graphicx}
\usepackage{epsfig}
\usepackage{graphics}
\usepackage{amsmath}
\date{\today}
\pacs{number}
\begin{document}
%%%%%%%%%%%%%%%%%%%%%%%%%%%%%%%%%%%%%%%%%%%%%%%%%%%%%%%%%%%%%%%%%%%%%%%%%%%%%%%%%%%%%%%%%%%%%%%%%%%%%%%%%%%%%%
\title{Klein-Gordon Oscillator in Kaluza-Klein Theory}
%%%%%%%%%%%%%%%%%%%%%%%%%%%%%%%%%%%%%%%%%%%%%%%%%%%%%%%%%%%%%%%%%%%%%%%%%%%%%%%%%%%%%%%%%%%%%%%%%%%%%%%%%%%%%%
\author{Josevi Carvalho}\email{josevi@ccta.ufcg.edu.br}
\affiliation{Unidade Acad\^emica de Tecnologia de Alimentos, Centro de Ci\^encias e Tecnologia Agroalimentar, 
Universidade Federal de Campina Grande, Pereiros, Pombal, PB, 58840-000, Brazil}

\author{Alexandre M. de M. Carvalho}\email{alexandre@fis.ufal.br}
\affiliation{Instituto de F\'{\i}sica, Universidade Federal de Alagoas, Campus A. C. Sim\~oes - Av. Lourival Melo Mota, s/n, Tabuleiro do Martins - Macei\'o - AL, CEP: 57072-970}
\author{Everton Cavalcante}
\email{everton@ccea.uepb.edu.br}
\affiliation{Departamento de F\'isica, Universidade Federal da Para\'iba, Caixa Postal 5008, 58051-970, Jo\~ao Pessoa, PB, Brazil;\\ Centro de Ci\^encias
Exatas e Sociais Aplicadas, Universidade Estadual da Para\'{\i}ba, Patos, PB, Brazil}

\author{Claudio Furtado}\email{furtado@fisica.ufpb.br}
\affiliation{Departamento de F\'{\i}sica, CCEN,  Universidade Federal 
da Para\'{\i}ba, Cidade Universit\'{a}ria, 58051-970 Jo\~ao Pessoa, PB,
Brazil}

%%%%%%%%%%%%%%%%%%%%%%%%%%%%%%%%%%%%%%%%%%%%%%%%%%%%%%%%%%%%%%%%%%%%%%%%%%%%%%%%%%%%%%%%%%%%%%%%%%%%%%%%%%%%%%

%%%%%%%%%%%%%%%%%%%%%%%%%%%%%%%%%%%%%%%%%%%%%%%%%%%%%%%%%%%%%%%%%%%%%%%%%%%%%%%%%%%%%%%%%%%%%%%%%%%%%%%%%%%%%%
\begin{abstract}
In this contribution we study the Klein-Gordon oscillator on the curved background within the Kaluza-Klein theory. The problem of interaction between particles coupled harmonically with  topological defects in Kaluza-Klein theory is studied. We consider a series of topological defects, then treat the Klein-Gordon oscillator coupled   to this background and find the energy levels and corresponding eigenfunctions in these cases. We show that the energy levels depend on the global parameters  characterizing these spacetimes.  We also investigate a quantum particle  described by the Klein-Gordon oscillator interacting with a cosmic dislocation  in Som-Raychaudhuri spacetime in the presence  of homogeneous magnetic field in a Kaluza-Klein theory. In this case, the spectrum of energy is determined, and we  observe that these energy levels  represent themselves as the sum of the  term related with Aharonov-Bohm flux   and of  the parameter associated to the rotation of the spacetime. 
\end{abstract}
\maketitle

\section{Introduction}
\label{1}
Harmonic interactions play an important role in physics mainly when we  consider  the motion of particles in presence of molecular potentials and electromagnetic fields. 
Particularly, the harmonic oscillator  appears as a prototype model in many areas of physics such as solid states physics, quantum statistical mechanics and quantum field theory, 
and indeed serves as  an important physical  example to study the concepts and mathematical tools in standard quantum physics.  At the same time, within the relativistic 
quantum mechanics,  the  effects introduced by the peculiar motion of the particles   in physical system  at the high energy can be considered. Together, 
quantum and relativistic effects have received a  great attention  within the quantum treatment of dynamics of particles occurring in backgrounds produced by topological 
defects \cite{furtado,furtado1,furtado2,azevedo}.  A well known version for relativistic harmonic oscillator  was proposed in Ref~\cite{moshinsky} for spin-$1/2$ particle.  This oscillator was named f as Dirac oscillator. In the non-relativistic limit this model has  a behavior  of  harmonic oscillator with
a very strong spin-orbit coupling term.  This Dirac oscillator is characterized by a new coupling  of the momentum  of the particle that is linear in the coordinates. The most recently,  the relativistic  harmonic oscillator was  studied in a commutative
and noncommutative field theory \cite{moshinsky,mirza} among other configurations including magnetic fields \cite{vilalba,ferk,luis}. The Dirac oscillator was
investigated for spin-$1/2$ particle  in the presence of topological defects in Refs. \cite{josevipra,plaknut,epjpknut,knutaop,grgknut, edil1,edil2,edil3,edil4}. However,  these studies were carried out for quantum dynamics  of spin-1/2 particles, leaving a gap in treatment of harmonic interaction   for   relativistic  scalar particles. 

Several studies have  demonstrated an  interest  in relativistic  models \cite{20,21,22,23} where the interaction potential  is similar to that one of the harmonic oscillator, such as the 
vibrational spectrum of diatomic molecules \cite{24}, the binding of heavy quarks \cite{25,26} and the oscillations of atoms in crystal lattices, by mapping them as a position-dependent mass system \cite{27,28,29,30}. The importance of these potentials arises from the presence of a strong potential field. Recently, Bahar
and Yasuk \cite{20}, intending to study  the quark-antiquark interaction \cite{qqint,qqint2}, have investigated  a problem of a relativistic spin-0 particle possessing a position-dependent mass, where the mass term acquires a contribution given by  an interaction potential that consists of a linear and a harmonic confining  potentials plus a Coulomb
potential term.  

The Klein-Gordon oscillator \cite{bruce,dvo}  was inspired by earlier papers on  the Dirac oscillator \cite{moshinsky} applied to half-integer spin particles. Recently Rao {\it et al.} \cite{rao} 
studied the spectral distribution of energy levels and eigenfunctions sets  describing the state of a particle by solving the Klein-Gordon equation in one-dimensional version
of Minkowski spacetimes.  In  the recent paper, Boumali and  Messai \cite{boumali} have  investigated a Klein-Gordon oscillator in  the background of cosmic string in the presence of an uniform magnetic field. The Klein-Gordon oscillator was investigated in the presence of Coulomb potential considering two ways of  coupling of Coulomb  potential  within the Klein-Gordon equation: in the first  paper \cite{bakkefur} the  Coulomb potential is incorporated in the equation by a modification mass term, in the  second model \cite{bakkefur2} this potential is introduced in  the Klein-Gordon equation via  the minimal coupling, in this last case the linear potential also  was  included in the equation.  Our intention now is to  extend these  studies  not  only to other dimensions but  mostly to consider this dynamics in general background spacetimes
produced by topological defects using  the  Kaluza-Klein theory \cite{kaluza,azreg,klein, furtado1,cbmmpla,knutkk}.  These  sources of gravitational fields play an important role in Condensed Matter 
Physics systems \cite{moraes1,moraes2,moraes3,moraes4}, mainly due  to the possibility to compensate the elastic contribution introduced by the defect by  a  fine
tuning of external magnetic field. Besides  of  topological defects  like  cosmic strings \cite{vilenkin} and  domain walls \cite{vilenkin1},  a global monopole \cite{barriola}
provides a tiny relation between effects in cosmology and gravitation and those in  Condensed Matter Physics  systems, where topological defects analogous to cosmic strings appear in phase transitions  in liquid crystals  \cite{breno,moraes5}. Recently,  the Klein-Gordon oscillator in the  Som-Raychaudhuri spacetime in the presence of  an uniform magnetic fields is investigated by Wang {\it et al.} \cite{wangepjp}. 

This contribution is organized as  follows:  in the section \ref{sec2}, using the Kaluza-Klein theory, we  study the energy levels of  particles interacting with
gravitational field produced by  cosmic string in the presence of the Klein-Gordon oscillator.  In  the section \ref{sec3}, we study the quantum dynamics in the presence of the magnetic cosmic string, calculating the spectral energy as well as the  corresponding  eigenfunctions. In the section \ref{sec4}, we consider  the case in which the background  has a torsional source of  a gravitational field added to  the curvature source introduced by the conical defect. To consider the contribution introduced by a magnetic field to this dynamics in the section \ref{sec5}, we  consider a homogeneous magnetic field filling the space accessible to Klein-Gordon particle. Additionally to the magnetic field we consider  rotational  spacetimes, whose rotation is introduced via geometric description of Kaluza-Klein theory. In (\ref{summary}), some
 discussion of our results will be presented. Throughout the article we will consider  the system of unities where $\hbar=c=G=1$.

%%%%%%%%%%%%%%%%%%%%%%%%%%%%%%%%%%%%%%%%%%%%%%%%%%%%%%%%%%%%%%%%%%%%%%%%%%%%%%%%%%%%%%%%%%%%%
\section{Klein-Gordon Oscillator in Cosmic String Background}\label{sec2}
%%%%%%%%%%%%%%%%%%%%%%%%%%%%%%%%%%%%%%%%%%%%%%%%%%%%%%%%%%%%%%%%%%%%%%%%%%%%%%%%%%%%%%%%%%%%%

The purpose of this section is to study the Klein-Gordon oscillator in the background of the cosmic string  with use of the Kaluza-Klein theory~\cite{azreg,furtado1,kaluza,klein}.  A first study of  a topological defect in Kaluza-Klein theory was  carried out in Ref. \cite{azreg}, where  the authors have investigated  a series   of  cylindrically symmetric solutions of Einstein and Einstein-Gauss-Bonet equations. In \cite{azreg}, they have found various solutions of  a cosmic string  form in five dimension, such as:  neutral cosmic string, cosmic dislocation, superconducting cosmic, multi-cosmic string spacetime.   The metric corresponding to this geometry can be written as,
\begin{eqnarray}
 ds^2=-dt^2+d\rho^2+\alpha^2\rho^2d\phi^2+dz^2+dx^2
\end{eqnarray}
where $t$ is the time coordinate, $x$ is the coordinate associated with the fifth additional dimension and $(\rho,\phi,z)$ are cylindrical coordinates. These coordinates 
assume, respectively, the following range $-\infty<(t,z)<\infty$, $0\leq\rho<\infty$, $0\leq\phi\leq 2\pi, 0<x< 2\pi a$,  where $a$ is the radius of  the  compact dimension $x$. The $\alpha$ parameter  characterizing the cosmic string,
is given in terms of mass density of the string $\mu$, by $\alpha=1-4\mu$ \cite{vilenkin, alpha}. The cosmology and gravitation imposes limits to the range of the $\alpha$ parameter which 
is restricted to $\alpha<1$ \cite{vilenkin}. Moreover, in Condensed Matter Physics systems, this restriction is free and the opposite case $\alpha>1$,  the known negative disclination  \cite{katanaev}, 
can occur in several systems as those described by~\cite{moraes5}.

To  couple the Klein-Gordon oscillator \cite{bruce,dvo}  to this background we use the  generalization of   Mirza and Mohadesi prescription \cite{mirza}, in which, we carry out the following change in momentum  operator:
\begin{eqnarray}
p_{\mu} \to (p_{\mu} + iMw X_{\mu})
\end{eqnarray}
where we have defined in polar  coordinates $X_{\mu}=(0,\rho,0,0,0)$,  $\rho$ is the transverse  distance  from the particle to the defect.  In this way, the general Klein-Gordon equation  becomes
\begin{eqnarray}\label{klein}
\left\{\frac{1}{\sqrt{-g}}(\partial_{\mu}+MwX_{\mu})\sqrt{-g}g^{\mu\nu}(\partial_{\nu}-MwX_{\nu})-M^2\right\}\Psi=0
\end{eqnarray}
with $g$ being the determinant of the metric tensor and $g^{\mu\nu}$ is the inverse  metric tensor, $\partial_{\mu,\nu}$ are  partial derivatives with
respect to the  coordinates. The matrix $g^{\mu\nu}$ is  of the form
\begin{eqnarray}
g^{\mu\nu} = diag(-1,1,1/\alpha^2\rho^2,1,1),
\end{eqnarray}
in this way the equation (\ref{klein}) becomes
\begin{eqnarray}\label{mklein}
\left\{\frac{1}{\rho}\partial_\rho(\rho\partial_\rho)\right.&+&\left.\frac{\partial_\phi^2}
{\alpha^2\rho^2}-M^2w^2\rho^2+\gamma\right\}\Psi=0
\end{eqnarray}
with $\gamma=-\partial_t^2+\partial_z^2+\partial_x^2-2Mw-M^2$ and $x$ is fifth spatial coordinate in Kaluza-Klein theory. Supposing a temporal independence  of
our background and  translational symmetry along the axis {\it x} and {\it z}, we can  choose  the following {\it ansatz},
\begin{eqnarray}\label{unsatz}
 \Psi=e^{-i(Et-kz-l\phi-\lambda x)}R(\rho),
\end{eqnarray}
in this way, the equation (\ref{mklein}) transforms into
\begin{eqnarray}
 \left\{\frac{1}{\rho}\partial_\rho(\rho\partial_\rho)-\left[\frac{l^2}{\alpha^2\rho^2}+M^2w^2\rho^2-\gamma\right]\right\}R(\rho)=0,
\end{eqnarray}
that, with help of (\ref{unsatz}) the last term in above equation is  rewritten as $\gamma=E^2-M^2-2Mw-k^2-\lambda^2$. This equation can be transformed  into other 
one by the following change  of the variable $\xi=Mw\rho^2$. The result is
\begin{eqnarray}
\label{eqr}
 R''(\xi)+\frac{1}{\xi}R'(\xi)-\left(\frac{l^2}{4\alpha^2\xi^2}+\frac{1}{4}-\frac{\gamma}{4Mw\xi}\right)R(\xi)=0
\end{eqnarray}
 Now we proceed  with study of the asymptotic limit  at origin and infinity.  In what follows, we suppose,
\begin{eqnarray}
 R(\xi)=\xi^{\frac{|l|}{2\alpha}}e^{-\frac{\xi}{2}}F(\xi)
\end{eqnarray}
Substituting  of $R(\xi)$ in this form in the equation (\ref{eqr}) results in
\begin{eqnarray}\label{fhc}
\xi F''(\xi)+\left(\frac{|l|}{\alpha}+1-\xi\right)F'(\xi)-\left(\frac{l}{2\alpha}+\frac{1}{2}
-\frac{\gamma}{4Mw}\right)F(\xi)=0
\end{eqnarray}
where $\gamma$ is as before. We can see that this equation  is just that one for the confluent hypergeometric function, $xF''(x)+(c+1-x)F'(x)-aF(x)=0$, whose
solution   is a polynomial of  the degree $n$. Naturally there  is a convergence problem in this solution when $n$ tends to infinity. To avoid this divergence, we
can choose the independent term, last term, in equation (\ref{fhc})  to be equal to a non-negative number. Mathematically we have,
\begin{eqnarray}
 \frac{|l|}{2\alpha}+\frac{1}{2}-\frac{\gamma}{4Mw}=-n,
\end{eqnarray}
using the respective expressions to $\gamma$ and solving the resultant equation to $E$ we obtain the following eigenvalues  for this problem:
\begin{eqnarray}
E^2=M^2+4Mw\left(n+\frac{|l|}{2\alpha}+1\right)+k^2+\lambda^2
\end{eqnarray}
and the following eigenfunctions sets
\begin{eqnarray}
\Psi(\vec r,t)=C_{n,l}\;e^{-i(Et-kz-l\phi-\lambda x)} \rho^{\frac{|l|}{\alpha}}e^{-\frac{Mw}{2}\rho^2}F\left(-n,\frac{|l|}{\alpha}+1, Mw\rho^2\right).
\end{eqnarray}
Note the dependence on nonlocal parameters of the background  for energy levels as well as for the eigenfunctions is responsible in  breaking of degeneracy
of the energy levels due  to  the presence of the parameter $\alpha$. We observe that the present  result  is similar with  that one obtained by Boumali and Messai \cite{boumali} 
for  the Klein-Gordon oscillator in the background of  the cosmic string in  Einstein gravity.   Further,  in the weak oscillator  limit  $w\to 0$, our particles  behave like free particles. Moreover,  if flat spacetime limit $\alpha\longrightarrow1$  is taken
 the results in \cite{rao} are reproduced.

%%%%%%%%%%%%%%%%%%%%%%%%%%%%%%%%%%%%%%%%%%%%%%%%%%%%%%%%%%%%%%%%%%%%%%%%%%%%%%%%%%%%%%%%%%%%%
\section{Klein-Gordon Oscillator in the background of a Magnetic Cosmic String in a Kaluza-Klein Theory}\label{sec3}
%%%%%%%%%%%%%%%%%%%%%%%%%%%%%%%%%%%%%%%%%%%%%%%%%%%%%%%%%%%%%%%%%%%%%%%%%%%%%%%%%%%%%%%%%%%%%
Now,  let us consider the quantum dynamics of a particle moving in the magnetic cosmic string background. In the Kaluza-Klein theory \cite{furtado1,kaluza,klein}, the
corresponding metrics with a magnetic flux $\Phi$  passing  along the symmetry axis of the string assumes the following  form,
\begin{eqnarray}
 ds^2=-dt^2+d\rho^2+\alpha^2\rho^2d\phi^2+dz^2+\left(dx+\frac{\Phi}{2\pi}d\phi\right)^2
\end{eqnarray}
with cylindrical coordinates  are used. The quantum dynamics is described by the equation (\ref{klein}) with the following change in the inverse matrix tensor $g^{\mu\nu}$,
\begin{eqnarray}
g^{\mu\nu} = \left(
\begin{array}{ccccc}
-1 & 0 & 0 & 0&0\\
0& 1 & 0 & 0 & 0\\
0 & 0 & \frac{1}{\alpha^2\rho^2} & 0 & -\frac{\Phi}{2\pi\alpha^2\rho^2}\\
0 & 0 & 0 &1 &0\\
0 & 0 & -\frac{\Phi}{2\pi\alpha^2\rho^2}&0 &1+\frac{\Phi^2}{4\pi^2\alpha^2\rho^2}\\
\end{array}
\right).
\end{eqnarray}
in this way the equation (\ref{klein}) becomes,
\begin{eqnarray}\label{mklein1}
\left\{\frac{1}{\rho}\partial_\rho(\rho\partial_\rho)\right.&+&\left.\frac{[\partial_\phi-(\Phi/2\pi)\partial_x]^2}{\alpha^2\rho^2}-M^2w^2\rho^2+\gamma\right\}\Psi=0
\end{eqnarray}
with $\gamma=-\partial_t^2+\partial_z^2+\partial_x^2-2Mw-M^2$. Supposing a temporal independence  of our background, we can  choose the following ansatz
\begin{eqnarray}\label{unsatz1}
 \Psi=e^{-i(Et-kz-l\phi-\lambda x)}R(\rho)
\end{eqnarray}
thus, the equation (\ref{mklein1}) transforms into
\begin{eqnarray}
 \left\{\frac{1}{\rho}\partial_\rho(\rho\partial_\rho)-\left[\frac{(l-\lambda\Phi/2\pi)^2}{\alpha^2\rho^2}+M^2w^2\rho^2-\gamma\right]\right\}R(\rho)=0,
\end{eqnarray}
that with help of (\ref{unsatz1}) the last term in the above equation is $\gamma=E^2-M^2-2Mw-k^2-\lambda^2$.  After the change of variables $x=Mw\rho^2$, this equation takes the form
\begin{eqnarray}\label{gen1}
 R''(x)+\frac{1}{x}R'(x)-\left(\frac{\beta^2}{4\alpha^2x^2}+\frac{1}{4}-\frac{\gamma}{4Mwx}\right)R(x)=0
\end{eqnarray}
Now, we study the  behavior of our function in the  small $x$ and large $x$  limits. Asymptotically  at $x\to 0$ we find the following expression,
\begin{eqnarray}
 R''(x)+\frac{1}{x}R'(x)-\frac{\beta^2}{4\alpha^2x^2}R(x)=0,
\end{eqnarray}
whose solution  is given by  $R(x)=x^{\frac{|\beta|}{2\alpha}}$. In another  limit $x\to\infty$ the resultant equation is,
\begin{eqnarray}
 R''(x)-\frac{1}{4}R(x)=0,
\end{eqnarray}
 which yields $R(x)=e^{-\frac{x}{2}}$. A general solution can be determined by choosing
\begin{eqnarray}
 R(x)=x^{\frac{|\beta|}{2\alpha}}e^{-\frac{x}{2}}F(x).
\end{eqnarray}
 Substitution of  this expression into equation (\ref{gen1}) results in
\begin{eqnarray}\label{fhc1}
x F''(x)+\left(\frac{|\beta|}{\alpha}+1-x\right)F'(x)-\left(\frac{\beta}{2\alpha}+\frac{1}{2}-\frac{\gamma}{4Mw}\right)F(x)=0
\end{eqnarray}
where $\gamma$ is as before and $\beta=l-\lambda\Phi/2\pi$. Again this equation is in the  same form as  that one describing the confluent hypergeometric function $xF''(x)+(c+1-x)F'(x)-aF(x)=0$. 
Requiring the convergence of the corresponding series, we arrive  at the following condition
\begin{eqnarray}
 \frac{\beta}{2\alpha}+\frac{1}{2}-\frac{\gamma}{4Mw}=-n.
\end{eqnarray}
Using the respective expressions  for $\beta$ and $\gamma$ and solving the resultant equation for $E$, we obtain the following eigenvalues problems,
\begin{eqnarray}
E^2=M^2+4Mw\left(n+\frac{|l-\lambda\Phi/2\pi|}{2\alpha}+1\right)+k^2+\lambda^2
\end{eqnarray}
Comparatively to the case without magnetic flux string in the center  of  the defect, the angular quantum number is shifted by the quantity related with the magnetic field due  to a presence of Aharonov-Bohm flux in a magnetic string.  Additionally, we can  see that the presence of the defects breaks the degeneracy of the energy levels due to the presence of the curvature source. Besides, in absence of magnetic fields, $\Phi=0$, the  results of previous section are obtained. Note that the presence  of magnetic string and the Aharonov-Bohm flux,  modifies the energy spectrum, this effect is well known  as  Aharonov-Bohm effect for  a bound state \cite{cbmmpla,ahbound}, in fact the energy levels are shifted by a quantity proportional to Aharonov-Bohm flux. 

%%%%%%%%%%%%%%%%%%%%%%%%%%%%%%%%%%%%%%%%%%%%%%%%%%%%%%%%%%%%%%%%%%%%%%%%%%%%%%%%%%%%%%%%%%%%%
\section{Klein-Gordon oscillator in Cosmic Dispiration background in a Kaluza-Klein Theory}\label{sec4}
%%%%%%%%%%%%%%%%%%%%%%%%%%%%%%%%%%%%%%%%%%%%%%%%%%%%%%%%%%%%%%%%%%%%%%%%%%%%%%%%%%%%%%%%%%%%%
Now we investigate the Klein-Gordon oscillator in  a cosmic dispiration background in Kaluza-Klein theory \cite{furtado1}.  Let us  study the concurrency between gravitational  effects due to the torsion  and curvature, and 
electromagnetic  contributions  due to the presence of this topological defect. In this way, we consider the magnetic cosmic string with torsion source
besides  of  curvature and electromagnetic ones. The corresponding background is described by  the metric \cite{furtado1},
\begin{eqnarray}
ds^2=-dt^2+d\rho^2+\alpha^2\rho^2d\phi^2+(dz+Jd\phi)^2+\left(dx+\frac{\Phi}{2\pi}d\phi\right)^2
\end{eqnarray}
with $\alpha$, $J$ and $\Phi$, respectively, the sources of curvature, torsion and electromagnetic  field. The coordinates $(t,\rho,\phi,z,x)$ are defined as before, and the inverse
 metric tensor is
\begin{eqnarray}
g^{\mu\nu} = \left(
\begin{array}{ccccc}
-1 & 0 & 0 & 0 &0\\
0& 1 & 0 & 0 & 0\\
0 & 0 & \frac{1}{\alpha^2\rho^2} & -\frac{J^2}{\alpha^2\rho^2} & -\frac{\Phi}{2\pi\alpha^2\rho^2}\\
0 & 0 & \frac{J}{\alpha^2\rho^2} &1+\frac{J^2}{\alpha^2\rho^2} &\frac{\Phi\;J}{2\pi\alpha^2\rho^2}\\
0 & 0 & -\frac{\Phi}{2\pi\alpha^2\rho^2}& \frac{\Phi\;J}{2\pi\alpha^2\rho^2} &\left(1+\frac{\Phi^2}{4\pi^2\alpha^2\rho^2}\right)\\
\end{array}
\right).\quad
\end{eqnarray}

In this way the equation (\ref{klein}) becomes
\begin{eqnarray}\label{partial1}
\left\{\frac{\partial^2}{\partial\rho^2}+\frac{1}{\rho}\frac{\partial}{\partial\rho}+\frac{(\partial_\phi-J\partial_z-
\Phi\partial_y/2\pi)^2}{\alpha^2\rho^2}-M^2w^2\rho^2+\gamma\right\}\Psi(t,\vec r)=0,
\end{eqnarray}
with $\gamma=-\partial_t^2+\partial_z^2+\partial_x^2-2Mw-M^2$. This partial differential equation  does not  involve explicit dependence on variables $t,\phi,z$
and $x$, and therefore, has translational symmetry around these axes. These properties allow us  to suppose  a general solution  of Eq. (\ref{partial1}) in the
form $e^{-iEt+il\phi+ikz+i\lambda y}R(\rho)$. Thus,
\begin{eqnarray}\label{mklein2}
%\left\{ %\right.&-&\left. \right\}
\frac{1}{\rho}\frac{d}{d\rho}\left[\rho \frac{dR(\rho)}{d\rho}\right]-\left[\frac{\Lambda^2}{\alpha^2\rho^2}-M^2w^2\rho^2-\gamma\right]R(\rho)=0
\end{eqnarray}
Now $\Lambda = l-Jk-\lambda\Phi/2\pi$ and $\gamma=E^2-M^2-2Mw-k^2-\lambda^2$. Making $\xi = Mw\rho^2$, this radial equation transforms into
\begin{eqnarray}
 \frac{d^2R(\xi)}{d\xi^2}+\frac{1}{\xi}\frac{dR(\xi)}{d\xi}-\left(\frac{\Lambda^2}{4\alpha^2\xi^2}+\frac{1}{4}-\frac{\gamma}{4Mw\xi}\right)R(\xi)=0
\end{eqnarray}
An analysis of the divergence at  the origin and infinity suggest use of  the general solution $R(\xi) = \xi^\frac{|\Lambda|}{2\alpha}e^{-\frac{\xi}{2}}F(\xi)$. After  some calculations,
the final  equation  for $F(\xi)$, that have the same form as the confluent hypergeometric equation,
\begin{eqnarray}\label{final}
\xi\frac{d^2F(\xi)}{d\xi^2}&+&\left(c+1-\xi\right)\frac{dF(\xi)}{d\xi}-aF(\xi)=0.\nonumber\\ 
\xi\frac{d^2F(\xi)}{d\xi^2}&+&\left(\frac{|\Lambda|}{\alpha}+1-\xi\right)\frac{dF(\xi)}{d\xi}-\left(\frac{|\Lambda|}{2\alpha}+\frac{1}{2}-\frac{\gamma}{4Mw}\right)F(\xi)=0\nonumber\\
 &\Lambda& = l-Jk-\frac{\lambda\Phi}{2\pi}.
\end{eqnarray}

A solution for this equation is  a polynomial of degree $n$ in $\xi$ of the form $F(\xi) = \sum{a_n}\xi^n$. We can see that this solution  diverges for all
values of $n$  which represents the degree of the hypergeometric series. This divergence is  avoided by a truncation method in coefficients $a_n$. If we assume
that $a_n=0$  for a finite numbers of terms in the polynomial series, thus we guarantee normality of our solution in $\xi$ to $\xi\to 0$ and avoid the
divergence to $\xi\to\infty$. Making a general expression for the coefficients $a_n$, we conclude that to ensure this integrability, we  must make $a=-n$. Thus,
\begin{eqnarray}
\frac{|\Lambda|}{2\alpha}+\frac{1}{2}-\frac{\gamma}{4Mw} = -n
\end{eqnarray}
This condition give us the energy levels and eigenfunctions  for our scalar particle in the form
\begin{eqnarray}\label{eigendis}
E^2=M^2+4M\left(n+\frac{|l-Jk-\lambda\Phi/2\pi|}{2\alpha}+1\right)w+k^2+\lambda^2
\end{eqnarray}
and the following eigenfunctions sets
\begin{eqnarray}
\Psi(\vec r,t)=C_{n,l}\;e^{-i(Et-kz-l\phi-\lambda x)} \rho^{\frac{|\Lambda|}{\alpha}}e^{-\frac{Mw}{2}\rho^2}F\left(-n,\frac{|\Lambda|}{\alpha}+1, Mw\rho^2\right).
\end{eqnarray}
 We have now an important result: all global parameters of the background appear in energy levels of Klein-Gordon oscillator, however this background can be locally flat. The degeneracy of energy is absent due  to the presence of curvature and torsion sources. As  pointed out previously, this  result allows us to compensate the torsion contribution by adjusting the magnetic field appropriately, and the medium becomes torsion-free. Note that the term $\lambda\Phi/2\pi$ in (\ref{eigendis}) is responsible for electromagnetic Aharonov-Bohm effect \cite{cbmmpla,ahbound} for bound state due the presence of magnetic flux  $\Phi$ due  to a solenoid field in  the extra dimension. The term   $Jk$  in (\ref{eigendis}) is  present due  to a torsion of topological defect and $J$ is associated with the Burgers vector of cosmic dispiration, the quantum effect associated to this term in the energy spectrum of Klein-Gordon oscillator is responsible by the shift in this level well  known by gravitational Aharonov-Bohm  effect \cite{furtado1,furtado2} due to the torsion of this spacetime. In this way, we have three different contributions which modify the energy levels of the Klein-Gordon oscillator.  The first one is due to the conical nature of spacetime, represented by the deficit angle $\alpha$. The second  is due to  the contribution of torsion represented by $J$ and the third is due  to the electromagnetic field represented by $\Phi$, in  this form $E(\alpha,J,\Phi)$.

%%%%%%%%%%%%%%%%%%%%%%%%%%%%%%%%%%%%%%%%%%%%%%%%%%%%%%%%%%%%%%%%%%%%%%%%%%%%%%%%%%%%%%%%%%%%%%%%%%%%%%%%%%%%
\section{KG-Oscillator in Cosmic Dislocation in Som-Raychaudhuri spacetime in Kaluza-Klein Theory}\label{sec5}
%%%%%%%%%%%%%%%%%%%%%%%%%%%%%%%%%%%%%%%%%%%%%%%%%%%%%%%%%%%%%%%%%%%%%%%%%%%%%%%%%%%%%%%%%%%%%%%%%%%%%%%%%%%%
Recent observational data  indicate the rotation as well as  the expansion of our universe. This dynamics has called  a great attention to the establishment of a
theory  aimed to describe these scenarios. One  of  these theories was  developed  by G\"odel in the 1950  for an universe with rigid rotation characterized by a term
$\Omega$ in the metric and with  a curvature source  known as Weyssenhoff-Raabe fluid \cite{godel,som}. Some  studies of  quantum dynamics in this spacetime
were carried out for a (3+1)-dimensional spacetimes \cite{damiao,fiol,jcarvalho}. In this section we consider a cosmic dispiration in a flat G\"odel 
solution or Som-Raychaudhuri solution in Kaluza-Klein theory. We consider that  the charged scalar particle is  exposed to an uniform magnetic field. This 
field  is  also introduced using the Kaluza-Klein theory via  the geometry of the spacetime.  Due to the importance of the rotation in actual scenarios, we can use 
the Kaluza-Klein theory to describe  the quantum dynamics of a Klein-Gordon particle in this spinning background. Now,  we present a new solution for Som-Raychaudhuri spacetime  with  a topological defect  of a cosmic dispiration type, localized parallel to rotation axis.  We have considered this  solution in Kaluza-klein theory and have introduced  this  by extra dimensions, an  Aharonov-Bohm flux and an uniform magnetic field. In contrast with the  previous section where we have studied  quantum dynamics in  a topological defect background,  here we consider the influence of introduction of a cosmic string in G\"odel-type universe. Let us consider the Som-Raychaudhuri solution of the 
Einstein field equation   \cite{som} with a cosmic dispiration,  described in a Kaluza-Klein theory with a  spinning, torsion source along the
symmetry axis of background spacetimes,
\begin{eqnarray}
\label{som}
ds^2&=&-(dt+\alpha\Omega\rho^2d\phi)^2+d\rho^2+\alpha^2\rho^2d\phi^2+(dz+Jd\phi)^2+\nonumber\\&+&
\left[dx+(\Phi/2\pi+eB\rho^2/2)d\phi\right]^2.
\end{eqnarray}
Here $\Omega$ is associated with the rotational source of the space. This solution is  a backbone of a cosmology  occurring at a large scale.  The rotation parameter $\Omega$  characterizes this scale, this parameter also can viewed with as the scale of magnetic-type, or twist,
gravitational field \cite{fiol,das}.  This spacetime is characterized by the following causality safe region \cite{reboucas1,reboucas2} $0<\rho<1/\Omega$. In this region of this spacetime we  have  no closed timelike curves CTC\cite{reboucas1,reboucas2}. Recently the quantum dynamics of scalar and spinorial particle have  been studied in this spacetime and similarities with quantum dynamics in the presence of a external magnetic field was observed in fourdimensional solution of Som-Raychaudhuri \cite{fiol,das}  and  in M-Theory in Ref.\cite{hikida}.  Basing in this similarity, we have obtained the solution (\ref{som})  in Kaluza-Klein theory where we have considered a  Som-Raychaudhuri solution with cosmic dispiration and a inclusion of an uniform magnetic field via extra dimension
It is easy  to write the matrix $g_{\mu\nu}(\vec r)$ and from it obtain $g^{\mu\nu}(\vec r)$ as below,
\begin{eqnarray}
g^{\mu\nu} = \left(
\begin{array}{ccccc}
\Omega^2\rho^2-1 & 0 & -\frac{\Omega}{\alpha} & \frac{\Omega\;J}{\alpha}&\left(\frac{\Omega\;\Phi}{2\pi\alpha}+\frac{eB\Omega}{2\alpha}\rho^2\right)\\
0& 1 & 0 & 0 & 0\\
-\frac{\Omega}{\alpha} & 0 & \frac{1}{\alpha^2\rho^2} & -\frac{J^2}{\alpha^2\rho^2} & -\left(\frac{\Phi}{2\pi\alpha^2\rho^2}+\frac{eB}{2\alpha^2}\right)\\
\frac{\Omega\;J}{\alpha} & 0 & \frac{J}{\alpha^2}{\rho^2} &1+\frac{J^2}{\alpha^2\rho^2} &\left(\frac{\Phi\;J}{2\pi\alpha^2\rho^2}+\frac{eBJ}{2\alpha^2}\right)\\
\left(\frac{\Omega\;\Phi}{2\pi\alpha}+\frac{eB\Omega}{2\alpha}\rho^2\right) & 0 & -\left(\frac{\Phi}{2\pi\alpha^2\rho^2}+\frac{eB}{2\alpha^2}\right)& 
\left(\frac{\Phi\;J}{2\pi\alpha^2\rho^2}+\frac{eBJ}{2\alpha^2}\right) &\left(1+\frac{\Phi^2}{4\pi^2\alpha^2\rho^2}+\frac{e\Phi\;B}{2\pi\alpha^2}+
\frac{e^2B^2\rho^2}{4\alpha^2}\right)\\
\end{array}
\right).\quad
\end{eqnarray}
In this way the square root of the determinant of this matrix  is given by $\sqrt{-g}=\alpha\rho$. Therefore we can write the Klein-Gordon  equation (\ref{klein}) in this background as
\begin{eqnarray}
& & \left\{\frac{\partial^2}{\partial\rho^2} + \frac{1}{\rho}\frac{\partial}{\partial\rho}+\frac{(\partial_\phi-J\partial_z-
\Phi\partial_y/2\pi)^2}{\alpha^2\rho^2}+\left[-M^2w^2+\left(\Omega\partial_t+\frac{eB\partial_y}{2\alpha}\right)^2\right]\rho^2+\gamma\right\}\Psi(t,\vec r)=0,\nonumber\\
\gamma &=& -\partial_t^2-\left(\partial_\phi-\;J\partial_z-\frac{\Phi\partial_y}{2\pi}\right)\left(\frac{2\Omega}{\alpha}\partial_t+\frac{eB}{\alpha^2}\partial_y\right)
+\partial_z^2+\partial_y^2-2Mw-M^2\quad
\end{eqnarray}
This equation is independent of the variables $t,\phi,z$ and $y$ and  allows us  to use an ansatz in the general form: $\Psi\propto e^{-iEt+il\phi+ikz+i\lambda y}R(\rho)$. Therefore,
\begin{eqnarray}
& &\frac{1}{\rho}\frac{d}{d\rho}\left[\rho\frac{d}{d\rho}R(\rho)\right]-\left\{\frac{(l-Jk-\lambda\Phi/2\pi)^2}{\alpha^2\rho^2}+\left[M^2w^2+\left(\frac{eB\lambda}{2\alpha}-
\Omega\;E\right)^2\right]\rho^2-\gamma\;\right\}R(\rho)=0;\nonumber\\
\gamma&=&E^2+\left(l-Jk-\frac{\lambda\Phi}{2\pi}\right)\left(\frac{eB\lambda}{\alpha^2}-\frac{2\Omega E}{\alpha}\right)-%\nonumber\\ &-& 
M^2-2Mw-k^2-\lambda^2.
\end{eqnarray}
It is not difficult to show that this equation is the same as a confluent hypergeometric equation  looking like
\begin{eqnarray}
 x\frac{d^2R(x)}{dx^2}&+&\left(\frac{|\beta|}{\alpha}+1-x\right)\frac{dR(x)}{dx}-\left(\frac{|\beta|}{2\alpha}+\frac{1}{2}-\frac{\gamma}{4\delta}\right)R(x)=0,\nonumber\\
 &\beta& = l-Jk-\frac{\lambda\Phi}{2\pi},\nonumber\\
 &\delta& = \sqrt{M^2w^2+\left(\frac{eB\lambda}{2\alpha}-\Omega\;E\right)^2}.
\end{eqnarray}
The solution of this equation is a polynomial of  $x^n$ order. For all limits, this function  diverges. This  blow-up is avoided assuming
$a=-n$ in  the general expression of the coefficients of the hypergeometric series. With this condition the energy levels of our particle are
\begin{eqnarray}\label{rotenergy}
 E^2=M^2&-&\left(l-Jk-\frac{\lambda\Phi}{2\pi}\right)\left(\frac{eB\lambda}{\alpha^2}-\frac{2\Omega E}{\alpha}\right)+
 4\sqrt{M^2w^2+\left(\Omega\;E-\frac{eB\lambda}{2\alpha}\right)^2}\times \nonumber\\&&\times\left(n+\frac{|l-Jk-\frac{\lambda\Phi}{2\pi}|}{2\alpha}
 +\frac{1}{2}\right)+2Mw+k^2+\lambda^2.
\end{eqnarray}
Apparently we can see that  in the absence of homogeneous magnetic field as well as  of rotation sources, the results of the previous section are  reproduced. We can  also see  that the external sources have  an important role in the dynamics of our particle due  to the explicit dependence on these parameters. Assuming that the physical
laws are true in any temporal scales, we believe that the rotation, magnetic fields, curvature and torsion sources had played an important role in the dynamics
of our universe in early evolution epoch and, therefore, our contribution has an interesting  application.

 It is important to study some limits.  Let us consider the weak oscillator as well as rotation- free limit of spacetimes,  which is equivalent to assume $(w , \Omega)\to 0$,
into equation (\ref{rotenergy}), in this limit we have  no influence  Klein-Gordon oscillator and the Som-Raychaudhuri geometry.  After  some calculations, we obtain the following result,
\begin{eqnarray}
E^2=M^2+\frac{2eB\lambda}{\alpha}\left[n+\frac{|l-kJ-\frac{\lambda\Phi}{2\pi}|}{2\alpha}-\frac{l-kJ-\frac{\lambda\Phi}{2\pi}}{2\alpha}+\frac{1}{2}\right]+k^2+\lambda^2,
\end{eqnarray}
that is exactly the result  of equation (35) of the Ref. \cite{furtado1}, where one of us have obtained the relativistic Landau levels for scalar particle in Kaluza-Klein theory.  Now we consider the limit where we do not have a Klein-Gordon oscillator, and obtain  from (\ref{rotenergy}) the following eigenvalues  of energy
\begin{eqnarray}\label{eigendisl}
E =&&\Omega \left(2n+\frac{|l-kJ-\frac{\lambda\Phi}{2\pi}|}{\alpha}-\frac{l-kJ-\frac{\lambda\Phi}{2\pi}}{\alpha}+1\right) \pm \nonumber \\
 && \left\{\left(\Omega^{2}\left(2n+\frac{|l-kJ-\frac{\lambda\Phi}{2\pi}|}{\alpha}-\frac{l-kJ-\frac{\lambda\Phi}{2\pi}}{\alpha}+\frac{1}{2}\right)-\frac{eB\lambda}{\alpha}\right)\right. \times \nonumber \\
&& \times\left.  \left[2n+\frac{|l-kJ-\frac{\lambda\Phi}{2\pi}|}{\alpha}-\frac{l-kJ-\frac{\lambda\Phi}{2\pi}}{\alpha} +1\right] +k^{2} + M^{2} + \lambda^{2}\right\}^{1/2}.
\end{eqnarray}
 These eigenvalues (\ref{eigendisl}) represent the energy levels for a free scalar particle in Som-Raychaudhuri  spacetime pierced by  a cosmic dispiration and an uniform magnetic field introduced in a geometric way by a Kaluza-Klein theory. Note the influence of topological defect in the eigenvalues: in the limit where we have $B\to 0$ in (\ref{eigendisl}) we obtain the eigenvalues of the quantum dynamics  of a scalar quantum particle in $(4+1)$-dimensional Som-Raychaudhuri  spacetime,  given by
\begin{eqnarray}\label{eigencosdis}
E =&&\Omega \left(2n+\frac{|l-kJ-\frac{\lambda\Phi}{2\pi}|}{\alpha}-\frac{l-kJ-\frac{\lambda\Phi}{2\pi}}{\alpha}+1\right) \pm \nonumber \\
 &\pm & \sqrt{\Omega^{2}\left(2n+\frac{|l-kJ-\frac{\lambda\Phi}{2\pi}|}{\alpha}-\frac{l-kJ-\frac{\lambda\Phi}{2\pi}}{\alpha}+\frac{1}{2}\right)^{2}   +k^{2} + M^{2} + \lambda^{2}}.
\end{eqnarray}
This results is the generalization of the results obtained previously in Refs. \cite{fiol,jcarvalho} for $(4+1)$-dimensional G\"odel  scenario in the presence of cosmic dispiration.  We can observe in (\ref{eigencosdis}) the dependence of the parameter $J$ that is related with the Burgers vector of  the topological defect  associated with the torsion of the spacetime.  
 In  the presence of rotation and curvature sources, one can consider the limit $(J,\Phi,B,w)\to 0$ 
in (\ref{rotenergy}). After a some algebra we obtain the result
\begin{eqnarray}
E=\left(2n+\frac{|l|}{\alpha}+\frac{l}{\alpha}+1\right)\Omega\pm\sqrt{\left(2n+\frac{|l|}{\alpha}+\frac{l}{\alpha}+1\right)^2\Omega^2+M^2+k^2}
\end{eqnarray}
that is equivalent to energy levels in the G\"odel-type spacetimes studied by us \cite{hikida,fiol,das,jcarvalho}. Although the Eq. (\ref{rotenergy})  has a cumbersome form, some 
limits  show important results in describing other  simple physical systems with topo\-logical/electromagnetic interactions through simple manipulations with that equation.
Finally, we can  see that the degeneracy of the energy levels, in this case, is strongly broken due to the presence of curvature, torsion, magnetic fields as well
as  of the rotation source.  It is important to observe a Landau structure in all previously studied cases.

%%%%%%%%%%%%%%%%%%%%%%%%%%%%%%%%%%%%%%%%%%%%%%%%%%%%%%%%%%%%%%%%%%%%%%%%%%%%%%%%%%%%%%%%%%%%%
\section{Summary}\label{summary}
%%%%%%%%%%%%%%%%%%%%%%%%%%%%%%%%%%%%%%%%%%%%%%%%%%%%%%%%%%%%%%%%%%%%%%%%%%%%%%%%%%%%%%%%%%%%%

The aim of this paper was  to investigate the quantum dynamics of  a scalar particle interacting harmonically  with gravitational background of topological defects, via Klein-Gordon oscillator  description, in  the presence of class of spacetimes in Kaluza-Klein theory. We determine the  manner in which the non-trivial topology due  to the topological defect, electromagnetic field and rotation of this background modify the energy spectrum and wave function of the Klein-Gordon oscillator. This  perturbation in the eigenvalues is compared with the flat spacetime, and these results can be used to investigated the presence of these defects in the cosmos. Here we investigate a harmonic interaction that can used for simulation of a series of  physical systems, such as,  vibrational spectrum of diatomic molecule \cite{24}, the binding of heavy quarks \cite{25,26}, quark-antiquark interaction\cite{qqint}. The possibility to use the  modification in the spectra of  KG-Oscillator  to probe the existence of this topological defects  was noticed in the obtained results. In fact  it is clear, from the observational point of view,   that to have a observable modification in the eigenvalues of energy, we need a huge number of  particles  in the states, otherwise the magnitude of effect to real spectrum  may not be strong enough to be observed.

 We have studied the quantum dynamics of a Klein-Gordon particle interacting with external  field sources, by using the five-dimensional version of the General Relativity. The quantum dynamics in the usual as well as magnetic cosmic string  cases allow us to obtain the energy levels and the eigenfunctions depending on the external parameters characterizing the background spacetimes, a result known by gravitational  analogue of the well studied Aharonov-Bohm effect.

We have investigated the Klein-Gordon oscillator in the cosmic string  background in a Kaluza-Klein theory and obtained the eigenvalues and  eigenfunctions of energy which  turn out to depend on the $\alpha$ parameter that  characterizes the cosmic string. Note that in the four-dimensional limit we  recover the results found by Boumali {\it et al.} \cite{boumali}.  For the case of Klein-Gordon oscillator in the presence of  the magnetic flux string in Kaluza-Klein theory, we  obtained that the energy levels  depend on the Aharonov-Bohm flux and the parameter $\alpha$.  The Klein Gordon equation for KG oscillator  in cosmic dispiration was investigated, and the energy levels  turn out to depend on $\alpha$ parameter,  the dislocation parameter (Burgers vector modulus) $J$ and the Aharonov-Bohm flux.  The torsion inclusion  has a important role in this dynamics. By the results in this background,  it becomes possible to compensate the elastic contribution introduced by the topological defect, by a fine tuning of the external magnetic field strength. The degeneracy of the energy levels  is strongly broken due  to the presence of curvature and torsional sources in these expressions.

Note that in the  sections \ref{sec2},\ref{sec3} and \ref{sec4} we have studied the influence of topological defect in Kaluza-Klein theory in energy levels of  a Klein-Gordon oscillator in order to observe the influence  of this structure in energy levels and  a wave function. In section \ref{sec5}, we have studied the Klein-Gordon oscillator in $(4+1)$-dimensional Som-Raychaudhury solution with a topological  defect, this study can be employed to investigate other quantum systems.

 We have obtained the spectrum and wavefunction for Klein-Gordon oscillator in  the background of the cosmic dispiration in a Som-Raychaudhury spacetime in   a Kaluza-Klein  theory in the presence of a uniform magnetic field and a magnetic flux. We introduced an uniform magnetic field and Aharonov-Bohm flux via Kaluza-Klein theory. The energy levels and  eigenfunctions  for Klein-Gordon oscillator in this geometry  were obtained, and we  demonstrated their dependence on the parameters characterizing the spacetime in $(4+1)$-dimension, such as , $(\Omega , \alpha , J, )$ associated  to the rotation, deficit angle and torsion of spacetime, the external magnetic field $B$ and the magnetic flux $\Phi$ introduced by Kaluza-Klein theory. Note that in an appropriate limit we obtain the results of previous section $\Omega\longmapsto0$, $B\longmapsto 0$,  that is cosmic dispiration case. In the case  where $\alpha=1$, $\Phi=0$ we obtain the results  to the ones of of Wang and  collaborators \cite{wangepjp}. In the limit of $\Omega=0$ the similar spectrum  of the Landau levels  is recovered.  Note that this dynamics  in the presence of the  confining potential  due to the Klein-Gordon oscillator   reinforces the characteristic of the quantum dynamics observed in this  G\"odel-type spacetime~\cite{fiol,das,jcarvalho},  which are characterized by a similarity of a Landau problem for a charged particle on a surface  exposed to  an uniform magnetic field.  Basing in this analogy,  Druker {\it et al. } \cite{fiol} have suggested a picture   of  a holographic description  for  a single chronologically safe region.  In  the same paper have conjectured that this discussion can be extended for $4+1$-G\"odel solution.  In this article we demonstrated that this similarity  with Landau levels occurs in $4+1$-G\"odel-type solution as well and have considered a more rich structure of  a spacetime including a topological defect. If we consider the analogy between the quantum dynamics in  chronologically safe region in Som-Raychaudhury spacetime and Landau levels,  we can  consider applications of these results of quantum dynamics in Hall droplets  of finite size \cite{halp}.  Because of this analogy, we can think of applications in Hall effect in  a  droplet of finite size in   systems of condensed matter. We can also use  the results found here in our harmonic confinement via Klein-Gordon oscillator in  Hall effect in droplets  of finite size with harmonic confinement of electrons, as   it was done in Ref. \cite{poly}.  We claim that these results can be used in a generalization of quantum Hall effect in $(4+1)$-dimensions \cite{pol,nair,nair2}, can be related with quantum dynamics in  gravity-based systems in higher dimension  due to the already mentioned analogy of quantum dynamics in G\"odel-type  solutions with Landau levels on curved surfaces \cite{comtet,dunne}.  In the  limit $w \to 0$ we have obtained  for  the first time the spectrum of  a scalar particle in Som-Raychaudhuri space time with a topological defect,  which combines the  curvature and the torsion into a dispiration, in the presence of a uniform field and magnetic flux introduced via Kaluza-Klein theory.  These energy levels (\ref{eigendisl}) have several contribution due the rotation of spacetime $\Omega$ of the parameter $J$ that is related with torsion.

%%%%%%%%%%%%%%%%%%%%%%%%%%%%%%%%%%%%%%%%%%%%%%%%%%%%%%%%%%%%%%%%%%%%%%%%%%%%%%%%%%%%%%%%%%%%%
{\bf Acknowledgment} \quad We  thank CAPES,  CNPQ, FAPESQ-PB for financial support.
%%%%%%%%%%%%%%%%%%%%%%%%%%%%%%%%%%%%%%%%%%%%%%%%%%%%%%%%%%%%%%%%%%%%%%%%%%%%%%%%%%%%%%%%%%%%%


\begin{thebibliography}{}
\bibitem{alpha} The parameter $\alpha= 1-4G\mu$ is adimensional,  in the present paper we have considered $ G=1$, in this way $\alpha= 1-4\mu$ \cite{vilenkin}.
\end{thebibliography}
\end{document}